\documentclass[english,preprint,onecolumn,showpacs,preprintnumbers,amsmath,amssymb,showkeys]{revtex4-2}

\usepackage{chemformula} 
\usepackage[T1]{fontenc} 
\usepackage{bm}
\usepackage[normalem]{ulem}
\usepackage[T1]{fontenc}
\usepackage{color}
\usepackage{units}
\usepackage{amssymb}
\usepackage{graphicx}
\usepackage{esint}
\usepackage{bm}
\usepackage{natbib}
\usepackage{ulem}
\usepackage{babel}
\usepackage[colorlinks=true, citecolor=red]{hyperref}

\begin{document}

	\title{Controlling Umklapp scattering in bilayer graphene moir\'{e} superlattice}
	\author{Mohit Kumar Jat$^1$, Shubhankar Mishra$^1$, Harsimran Kaur Mann$^1$, Robin Bajaj$^1$,  Kenji Watanabe$^2$, Takashi Taniguchi$^3$, H. R. Krishnamurthy$^1$, Manish Jain$^1$}
	\author{Aveek Bid$^1$}
	\email{aveek@iisc.ac.in}
	\affiliation{$^1$Department of Physics, Indian Institute of Science, Bangalore 560012, India \\
		$^2$ Research Center for Functional Materials, National Institute for Materials Science, 1-1 Namiki, Tsukuba 305-0044, Japan \\
		$^3$ International Center for Materials Nanoarchitectonics, National Institute for Materials Science, 1-1 Namiki, Tsukuba 305-0044, Japan\\}
	
	\setlength {\marginparwidth }{2cm}
	

	\begin{abstract}
	We present experimental findings on electron-electron scattering in two-dimensional moir\'e heterostructures with tunable Fermi wave vector, reciprocal lattice vector, and band gap. We achieve this in high-mobility aligned heterostructures of bilayer graphene (BLG) and hBN. Around half-filling, the primary contribution to the resistance of these devices arises from electron-electron Umklapp (Uee) scattering, making the resistance of graphene/hBN moir\'e devices significantly larger than that of non-aligned devices (where Uee is forbidden). We find that the strength of Uee scattering follows a universal scaling with Fermi energy and has a non-monotonic dependence on superlattice period. The Uee scattering is electric field tunable and is affected by layer-polarization of BLG. It has a strong particle-hole asymmetry -- the resistance when the chemical potential is in the conduction band is significantly lesser than when it is in the valence band, making the electron-doped regime more practical for potential applications.
	
\end{abstract}
\maketitle
\section{Introduction}
In a Galilean-invariant electron liquid, normal electron–electron scattering does not cause a loss of the momentum imparted to the electrons by the driving electric field; consequently, it can not lead to electrical resistance. A realistic Fermi liquid is, however, not Galilean invariant -- a finite coupling to an underlying lattice provides a mechanism for the momentum relaxation of the quasiparticles via the Umklapp process~\cite{Ashcroft76}. Umklapp electron-electron (Uee) scattering is the fundamental mechanism that allows momentum transfer from electrons to lattice and imparts electrical resistance to the metal~\cite{Kaveh1984ElectronelectronSI, Hideaki, Wallbank2019, PhysRevB.107.144111, Ishizuka_2022, RevModPhys.62.645,doi:10.1080/00018739300101514}. In this process, the crystal lattice gives a momentum kick to a pair of interacting electrons, backscattering them to the other side of the Fermi surface. Their quasi-momentum is conserved, modulo a reciprocal lattice vector $\mathbf{G}$,
\begin{eqnarray}
	\mathbf{k_1 + k_2 = k_3 + k_4 + G} \label{Eqn:Uee}
\end{eqnarray}
Here $\hbar \mathbf{k_{1,2}}$ and $\hbar \mathbf{ k_{3,4}}$ are the initial and final quasi-momenta of the two electrons near the Fermi level, respectively, and $\mathbf{G}$ is a non-zero reciprocal lattice vector of the crystal. This stringent conservation constraint, coupled with the lack of tunability of the Fermi wave vector, makes experimental identification of Uee processes in normal metals challenging~\cite{RevModPhys.62.645, PhysRevLett.78.705,doi:10.1080/00018739300101514}. Notable exceptions are heavy-fermionic systems whose large effective quasiparticle mass leads to an appreciable Uee-mediated resistance at very low temperatures ($\approx 100$ m$\mathrm{K}$)~\cite{Fisk1986}.

In the limit of nearly free electrons, one can view the Uee scattering as a two-stage process: In the first step, an electron-hole pair is excited into a virtual state by an electron, followed by the scattering of one of these particles by the periodic lattice potential. The temperature dependence of the Uee scattering process at a finite temperature is thus set by the size of the scattering phase space ($\propto k_B T/E_F$) for each electron; only the quasiparticles residing within a width of order $k_BT$ around the Fermi energy $E_F$ can undergo binary collisions. Consequently, the Uee contribution to the sheet resistance in $\mathrm{2D}$ goes as $\mathrm{\mathrm{R_\square}}_{Uee} = f_nT^2$~\cite{PhysRevB.23.2718}. $f_n \propto E_F^{-2}$ is a material-dependent parameter~\cite{https://doi.org/10.1002/andp.202100588,doi:10.1126/science.aaa8655, Wang2020}.

Note, however, that Uee need not be the only source of $T^2$ -resistivity in a material~\cite{PhysRevB.84.205111, doi:10.1126/science.aaa8655, PhysRevLett.93.267001, Maslov_2017, PhysRevB.4.2392}. A claim that the dominant source of scattering is the Uee process should be backed up by a (1) quantification of the prefactor $f_n$, (2) a demonstration of the scaling of $f_n \propto E_F^{-2}$, and (3) ruling out other competing mechanisms (e.g. electron-phonon scattering~\cite{PhysRevB.4.2392}) that can give $T$-dependent charge scattering.

Graphene-based moir\'e superlattices ~\cite{jat2023higherorder,Ishizuka_2022, Ponomarenko2013, PhysRevB.103.115419, ref1, Finney2019, doi:10.1126/sciadv.abd3655, Yankowitz2012, Wallbank2019} provide a system with precise tunability of the reciprocal lattice vectors $\mathbf{G}$ (via the twist angle between the constituent layers) and the Fermi wave vectors $\mathbf{k_F}$ (by controlling the carrier density $n$ through electrostatic gating). It thus provides a vast phase space in which Eqn.~\ref{Eqn:Uee} may be satisfied, and the scaling of $f_n$ versus $E_F$ can be verified. Recent calculations (that treat both the electron–electron Coulomb interaction and the moiré superlattice potential perturbatively) predict that in aligned heterostructures of Bernal bilayer graphene (BLG) and hBN, Uee scattering processes should be the primary source of resistance~\cite{PhysRevB.107.144111}.

In this Letter, we experimentally verify that in high-mobility moir\'e superlattices of BLG and hBN, Uee is the dominant source of resistance near half-filling. Our studies show that the strength of Uee depends non-monotonically on the superlattice period. This is at par with recent theoretical predictions~\cite{PhysRevB.107.144111} and in sharp contrast to observations in single-layer graphene-based superlattices~\cite{Wallbank2019}. We illustrate the tunability of strength of the Uee process (quantified by $f_n$) with displacement field, $D$ and carrier density, $n$. Additionally, we demonstrate a strong particle-hole asymmetry in the strength of the Uee process, whose origin can be traced to the moir\'{e} potential having a much stronger effect on the valence band than on the conduction band ~\cite{Wallbank2019, PhysRevB.107.144111}. Furthermore, we demonstrate the high tunability of Umklapp resistivity with an external vertical electric field, emphasizing the potential for precise control over the electronic properties of bilayer graphene superlattices. Finally, we show that these processes are completely absent in non-aligned devices.

\section{Results and discussion}
\begin{figure*}[t]
	\includegraphics[width=\columnwidth]{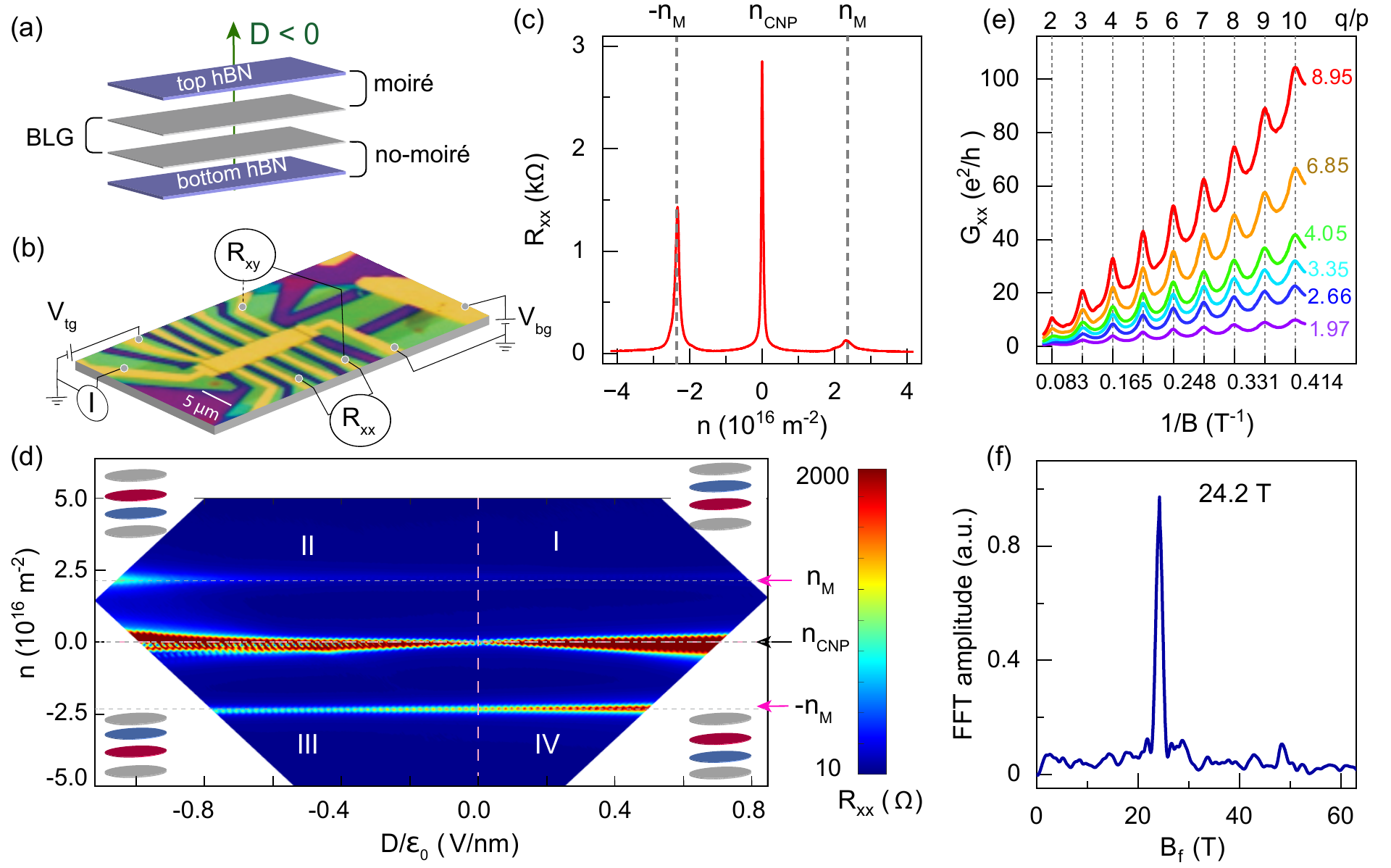}
	\small{\caption{\textbf{Characteristics of the moir\'{e} device M1.} (a) Schematic of the device layers, indicating moir\'{e} (no-moir\'{e}) superlattice formation between top hBN (bottom hBN) and the BLG. (b) An optical image of the device labeled with the measurement configuration (scale bar: $\mathrm{5 \mu m}$). (c) Plot of the longitudinal resistance $\mathrm{R_{xx}(B=0)}$ at $T=2$~$\mathrm{K}$ as a function of  $n$. Dotted gray lines mark the moir\'{e} satellite peaks with carrier density $n_{M} = \pm 2.30\times 10^{16}~\mathrm{m^{-2}}$. (d) $\mathrm{2D}$ map of $\mathrm{R_{xx}}$ as a function of $n$ and $D$. Labels I-IV mark the four quadrants in the $n-D$ plane. The four insets show schematically the charge distribution in the two layers of BLG in these four regimes at high $D$. The red (blue) ovals indicate the layers of BLG with the higher (lower) occupation of the electronic states. The upper bound on $\mathrm{R_{xx}}$ is set to be $2$~$\mathrm{k}$$\Omega$ for better visibility of the satellite peak (for the complete data, see Supplementary Information, section S2). (e) Plot of Brown-Zak oscillations $G_{xx}$ versus $1/B$ for different carrier  densities (units of $ 10^{16}~\mathrm{m^{-2}}$) measured at $T = 100$~$\mathrm{K}$. (f) The Fourier spectrum of the Brown-Zak oscillations measured at $n=4.05\times10^{16}$~$ \mathrm{m^{-2}}$ shows a single prominent peak at $B_f =24.2$~T. }
		
		\label{fig:fig1}}
\end{figure*}

High-quality hBN/BLG/hBN heterostructures were fabricated using the dry transfer technique (Supplementary Information, section S1)~\cite{doi:10.1021/acs.nanolett.3c00045, PhysRevLett.129.186802, Tiwari2022}. The top hBN was aligned at nearly zero degrees with BLG, and the bottom hBN was intentionally misaligned to a large angle to ensure that a moir\'{e} pattern forms only between top hBN and BLG (Fig.\ref{fig:fig1}(a)). The device is in hall bar geometry (Fig.\ref{fig:fig1}(b)) with dual gates to tune the carrier density $n$ and the vertical displacement field $D$ independently via $n=[(C_{tg}V_{tg}+C_{bg}V_{bg})/e+n_{r}]$ and $D=[(C_{tg}V_{tg}-C_{bg}V_{bg})/{2}+D_{r}]$. Here $C_{bg}$ ($C_{tg}$) is the back-gate (top-gate) capacitance, and $V_{bg}$ ($V_{tg}$) is the back-gate (top-gate) voltage.  $n_{r}$ and $D_{r}$ are the residual number density and displacement field in the graphene due to impurities. The direction of the negative displacement field ($D$) is marked schematically in Fig.\ref{fig:fig1}(a). In the main text, we provide the data for a device {M1} (with twist angle~$\approx 0^\circ$ and superlattice wavelength $~14$ nm), unless otherwise mentioned. The data for three more hBN/BLG/hBN superlattice devices, labeled {M2}, {M3} and {M4} with twist angle~$\approx 0.26^\circ, 0.47^\circ ~$and $~ 1.70^\circ$ respectively, and with superlattice wavelength ~$\approx13.64$~nm, $12.73$~nm and $7.20$~nm respectively, are presented in Supplementary Information. We also present data for a non-aligned hBN/BLG/hBN device (labeled {N1}) to compare the $T$-dependence of resistance between Uee-allowed (aligned devices) and Uee-forbidden (non-aligned devices) systems.

The measured longitudinal resistance $\mathrm{R_{xx}}$ on device {M1}, at $2$~$\mathrm{K}$ temperature shows a peak at the charge neutrality point (CNP), $n_{CNP} = 0$ and moir\'{e} satellite peaks at  $n_M = \pm 2.30\times 10^{16}~\mathrm{m^{-2}}$ (Fig.~\ref{fig:fig1}(c)). The mobility at CNP is extracted to be $350,000~\mathrm{m^{2}V^{-1}s^{-1}}$. Quantum Hall measurements at a perpendicular magnetic field of $B=5$~T establish that both spin and valley degeneracies are lifted, indicating the high quality of the device (Supplementary Information, section S5); these measurements are used to calibrate the values of $C_{bg}$  and $C_{tg}$. The angle homogeneity of the device is ascertained by comparing the $\mathrm{R_{xx}}$ data measured in different configurations (Supplementary Information, section S2).

Our results for $\mathrm{R_{xx}}$ as a function of carrier density $n$ and electric field $D/\epsilon_{0}$, shown in Fig.\ref{fig:fig1}(d) ascertain that the values of the moir\'e gap in carrier density $n_M$ are independent of the applied electric field. The plot can be divided into four quadrants labeled I-IV. In quadrants I ($n>0$, $D/\epsilon_{0}>0$)  and III ($n<0$, $D/\epsilon_{0}<0$), at a finite $D$, the occupied electronic states near the Fermi energy are predominantly localized (marked with a red color oval) in the bottom layer of BLG (away from moir\'{e} interface) and are weakly localized (marked with a blue color oval) in the top layer of BLG (close to moir\'{e} interface). This leads to the suppression of moir\'{e} effects and low resistance value of satellite peak in these quadrants. The opposite effect is seen in quadrants II ($n>0$, $D/\epsilon_{0}<0$)  and IV ($n<0$, $D/\epsilon_{0}>0$), the occupied electronic states are predominantly localized in the top layer of BLG (close to moir\'{e} interface) leading to the enhancement of moir\'{e} effects and higher resistance of the satellite peaks in these quadrants. Later in this Letter, we explore the consequences of this displacement-field-induced layer polarization on Uee scattering.

The moir\'{e} periodicity of the system is estimated from Brown-Zak oscillation measurements at $T = 100$~$\mathrm{K}$ (Fig.\ref{fig:fig1}(e)). Thermal broadening smears out Landau oscillations at this elevated temperature, and only Bloch oscillations survive ~\cite{PhysRevB.14.2239, krishna2018high, Huber2022, jat2023higherorder}. A Fourier spectrum of the oscillations yields the inverse periodicity or the `frequency' of the oscillations to be $B_f=24.2$~T (Fig.\ref{fig:fig1}(f)). Observation of only a single frequency rules out the double alignment of the BLG with hBN ~\cite{jat2023higherorder,doi:10.1126/sciadv.aay8897}. Using the relation  $S={h}/{eB_f}$ ($S$ being the real space area of the moir\'{e} superlattice cell, $h$: Planck's constant, $e$: electronic charge), the moir\'{e} wavelength is calculated to be $\lambda=14$~nm and the carrier density corresponding to filling the bands just up to the moir\'{e} gaps is $4/S = 2.30\times 10^{16}~\mathrm{m^{-2}}$; the factor of $4$ arises from the two-fold spin-and valley-degeneracy of graphene. This value of carrier density matches $n_M$ exactly, validating the number density corresponding to the moir\'e gap obtained from zero-magnetic field $\mathrm{R_{xx}}$ measurements. The twist angle between BLG and hBN corresponding to this moir\'{e} wavelength is approximately $0^\circ$ indicating near-perfect alignment between the top hBN and the BLG.

\begin{figure}[t]
	\includegraphics[width=\columnwidth]{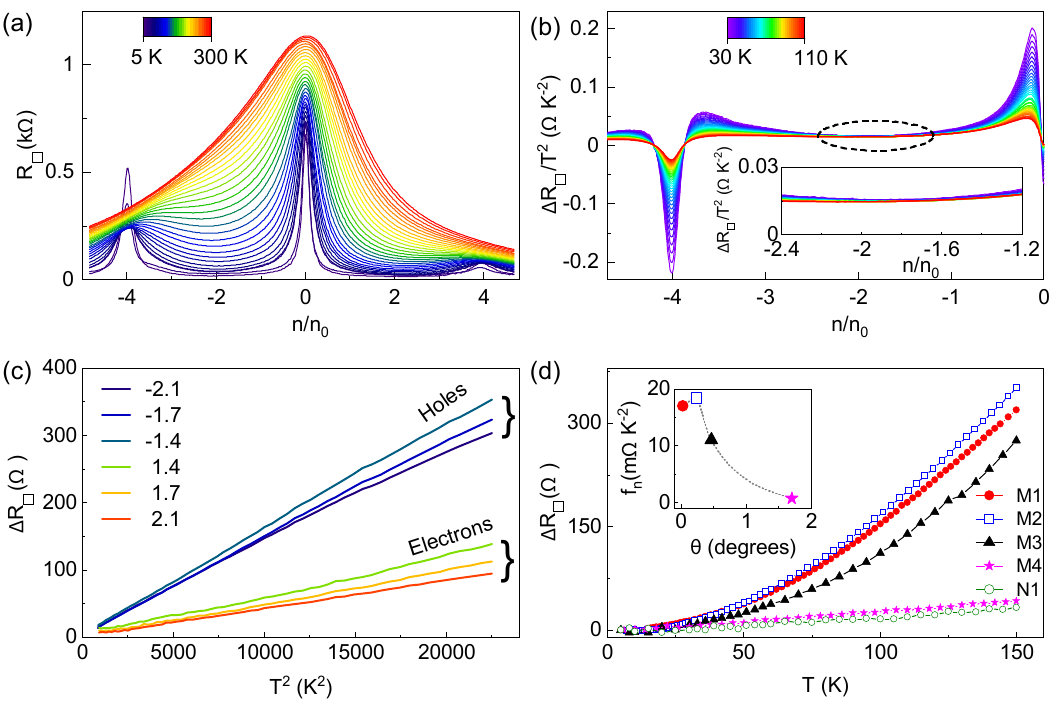}
	\small{\caption{\textbf{Umklapp scattering at $D/\epsilon_0 = 0$ V/nm.} (a) Plot of sheet resistance $R_{\square}$ as a function of filling fraction $n/n_0$ over a range of temperature from $5$~$\mathrm{K}$ (blue) to $300$~$\mathrm{K}$ (red). (b) Plot of $\Delta \mathrm{R_\square}/T^2= (\mathrm{R_\square}(T)-\mathrm{R_\square}(5K))/T^2$ versus $n/n_0$ over a range of temperature from $30$~$\mathrm{K}$ (blue) to $110$~$\mathrm{K}$ (red). The dotted ellipse marks the region where Umklapp is the dominant scattering mechanism. The negative value of $\Delta \mathrm{R_\square}/T^2$ around $n/n_0 = -4$ is a consequence of the fact that at these number densities, the value of $R_\square$ decreases with increasing $T$. Inset: Zoomed-in view of the region around $n/n_0=-2$. (c) Plot of $\Delta \mathrm{R_\square}$ as a function of $T^2$ for six different values of $n/n_0$. (d) Comparison of plots of $ \Delta \mathrm{R_\square}$ versus temperature at $n/n_0 = -2$ for four aligned devices ({M1}, {M2}, {M3} and {M4}) with twist angle ($0^\circ, 0.26^\circ, 0.47^\circ, 1.70^\circ$) and the non-aligned device {N1} at $n = -1\times 10^{16}~\mathrm{m^{-2}}$. Inset: Dependence of $f_n$ on the moir\'e twist angle $\theta$~(measured in degrees). The dashed line is a guide to the eyes.}
		
		\label{fig:fig2}}
\end{figure}

Fig.~\ref{fig:fig2}(a) shows the plots of the zero-magnetic field longitudinal sheet resistance $\mathrm{\mathrm{R_\square}} =\mathrm{R_{xx}}w/l$ ($w$ and $l$ are the width and length of the channel respectively with $w/l =1.5$) versus the moir\'e band filling fraction $n/n_0$ over a temperature range $5$~$\mathrm{K}$ $< T < 300$~$\mathrm{K}$ at zero displacement field. Here, $n_0 = 1/A = n_M/4$ is the carrier density at one-forth filling of the moir\'e band. With increasing temperature, one notices a sharp increase in $R_{\square}$  around $n /n_0=-2$ (Fig.\ref{fig:fig2}(d)); this feature is completely absent in non-aligned BLG devices (Supplementary Information, section S3). As we establish below, this rapid increase in ${\mathrm{R_\square}}$ with $T$ arises from the Umklapp scattering in the device.

At  $T = 0$, Uee is suppressed, and the resistivity is dominated by disorder scattering~\cite{https://doi.org/10.1002/andp.202100588}. To mitigate the effect of static disorder scattering, we henceforth focus on $\Delta \mathrm{\mathrm{R_\square}}(T) =\mathrm{\mathrm{R_\square}}(T)-\mathrm{\mathrm{R_\square}}(5~\mathrm{K})$. The magnitude of ${\mathrm{\mathrm{R_\square}}(5~\mathrm{K})}$  at $n/n_0=-2$ is $\approx~ 14\Omega$.
In  Fig.~\ref{fig:fig2}(b), we plot $\Delta \mathrm{\mathrm{R_\square}}/T^2$ versus $n/n_0$ over a temperature range from $30$~$\mathrm{K}$ to $110$~$\mathrm{K}$ -- the data at all temperatures collapse onto a single curve in the filling fraction range {$-2\leq n/n_0 \leq -1$}  (marked by the dotted ellipse) showing that $\Delta \mathrm{R_\square} \propto T^2$ over this range. This can be better appreciated from the inset, which shows the data over a narrow range around $n/n_0=-2$. Fig.~\ref{fig:fig2}(c) plots the $\Delta \mathrm{\mathrm{R_\square}}$ versus $T^2$ to better show the electron-hole asymmetry over a range of $n/n_0$. The linearity of the plots of sheet resistance versus $T^2$ in this carrier density regime persists till about  $T \simeq 110$~$\mathrm{K}$, establishing Uee scattering as the source of resistance (Fig.~\ref{fig:fig2}(c)). This temperature is of the order of the Bloch–Gr\"{u}neisen temperature in graphene. Above this $T$, electron-phonon scattering starts becoming the dominant source of resistance, and the quadratic relation between $\Delta \mathrm{R_\square}$ and $T$ breaks down~\cite{Nam2017, PhysRevB.91.121414, PhysRevB.83.235416}.
Fig.~\ref{fig:fig2}(d) shows a comparison of $\Delta \mathrm{R_\square}$ for the five devices -- the strong quadratic $T$-dependence seen in aligned devices is completely absent in the non-aligned device where Uee is forbidden from phase-space arguments. Inset in Fig.~\ref{fig:fig2}(d) shows $f_n$ versus twist angle for the aligned devices, illustrating the non-monotonicity of umklapp strength on the twist angle.

\begin{figure}[t]
	\includegraphics[width=\columnwidth]{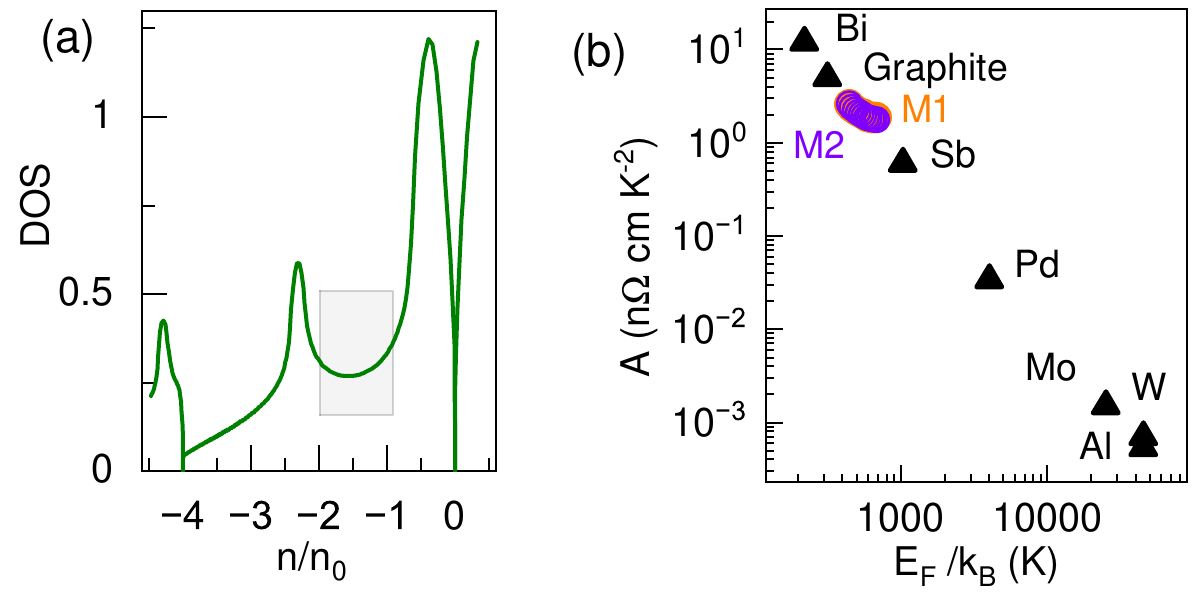}
	\small{\caption{\textbf{Universal scaling of Umklapp scattering.} (a) Plot of the calculated density of states (DOS) versus $n/n_0$. The shaded area marks the number density range, away from band edges and van Hove singularities, where Uee processes can be unambiguously detected. (b) Plot of $A= f_n t$  versus $E_F/k_B$. The open circles are the data from the current study on {M1} and {M2} superlattice devices. The filled triangles are the data from Ref~\cite{https://doi.org/10.1002/andp.202100588}.}
		
		\label{fig:fig3}}
\end{figure}

To understand the number density limits over which Umklapp processes are seen, recall that at very low $n/n_0$, transport in graphene is dominated by electron-hole puddles~\cite{Nam2017, Chen2008, PhysRevB.78.085415}; this gives a practical lower bound of $n/n_0$ at which e-e scattering is detectable~\cite{PhysRevB.107.144111}. A more accurate lower limit is obtained by the constraint that the Uee process imposes on the Fermi wave vector $\mathbf{k_F}$ (Eqn.~\ref{Eqn:Uee}), this sets a lower bound on $|n/n_0|$ equal to $  \pi /(2 \sqrt{3}) = 0.91$. (see Supplementary Information, section S4). At the other extreme, at high number densities, one begins to encounter electron-hole scattering processes at the principal mini band edges because of the moir\'{e} induced van-Hove singularity (Fig.\ref{fig:fig3}(b)), which masks the Umklapp scattering process~\cite{PhysRevB.107.144111}.

Before proceeding further, we eliminate the other probable causes that are known to lead to a $T^2$-dependence of the resistance. In a system with different carrier types/masses (as is the case near the primary and secondary gaps or van Hove singularities), the transfer of momentum between the two carrier reservoirs can lead to a resistivity with $T^2$ dependence~\cite{doi:10.1098/rspa.1937.0027, Xu2021, PhysRevB.53.11344}. This consideration guides us to avoid filling fractions that lead to Fermi levels close to these regions of the moir\'e bands and confine our analysis to the filling fraction range $-2\leq n/n_0 \leq -1$, as shown in~Fig.\ref{fig:fig3}(a). We note that, in low-mobility dilute alloys, the thermal motion of impurity ions can also give rise to a $T^2$-dependent resistance~\cite{10.1143/PTP.24.1049}; this scenario does not apply to our high-mobility heterostructures.

A phenomenological treatment, based on the Rice-Kadowaki–Woods scaling analysis~\cite{doi:10.1126/science.aaa8655,Wang2020} yields:
\begin{eqnarray}
	f_n \propto \frac{\hbar}{e^2}\left(\frac{k_B}{E_F}\right)^2
\end{eqnarray}

In  Fig.\ref{fig:fig3}(b), we plot $A= f_n t$  ($t  = 0.8$~nm is the thickness of BLG) as a function of the Fermi energy $E_F$ along with the compilation of data on several different materials~\cite{https://doi.org/10.1002/andp.202100588}. A very good match is obtained, emphasizing the universality of the value of $f_n$.

\begin{figure}[t]
	\includegraphics[width=\columnwidth]{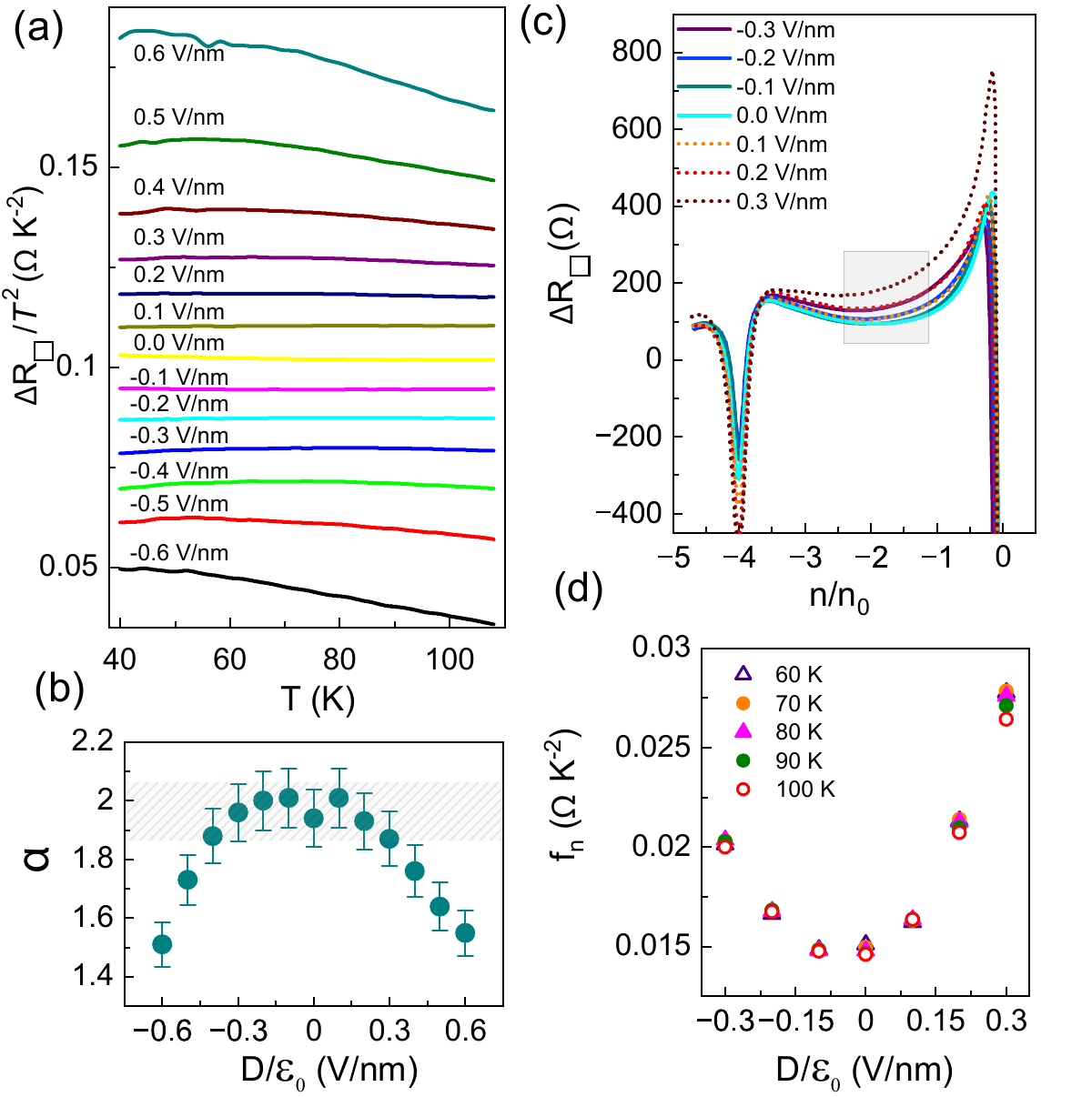}
	\small{\caption{\textbf{Electric field dependence of Umklapp scattering.} (a) Plots of $\Delta \mathrm{R_\square}/T^2$ versus $T$ for different values of $D/\epsilon_0$, the data are for  $n/n_0=-2 $. The numbers on the plots are values of the $D/\epsilon_0$. The data have been vertically offset for clarity. (b) Plot of the resistance exponent $\alpha$ ($\alpha = \mathrm{dln} (\Delta \mathrm{R_\square})/\mathrm{dln}(T)$) versus $D/\epsilon_0$ at $n/n_0 = -2$.  (c) Plot of $\Delta \mathrm{R_\square}$ versus filling fraction $n/n_0$ at temperature $T=80$~$\mathrm{K}$ for different values of $D/\epsilon_0$. (d) Plots of $f_n = \Delta \mathrm{R_\square}/T^2$ versus $D/\epsilon_0$ at a few representative values of $T$ in the Umklapp region at $n/n_0 = -2$.}
		
		\label{fig:fig4}}
\end{figure}

Having established Uee as the source of quasiparticle scattering in bilayer graphene/hBN moir\'e near half-filling ($n/n_0 =-2$), we now shift our focus on the effect of inter-layer potential asymmetry (tuned using $D$) on the Umklapp scattering in the quadrant III and IV of Fig.~\ref{fig:fig1}(d).    Fig.~\ref{fig:fig4}(a) plots $\mathrm{\Delta{R_\square}/T^2}(n/n_0=-2)$ versus $T$ for several different values of $D/\epsilon_{0}$. We find that the temperature exponent of the resistance $\alpha = \mathrm{dln} (\Delta \mathrm{R_\square})/\mathrm{dln}(T) \approx 2$ for $-0.3~\mathrm{V/nm} \leq D/\epsilon_{0} \leq 0.3~\mathrm{V/nm}$ (Fig.~\ref{fig:fig4}(b)). In this range of $D/\epsilon_{0}$, we find a substantial increase in the scattering strength with increasing $D/\epsilon_{0}$ in conformity with theoretical predictions~\cite{PhysRevB.107.144111} (Fig.~\ref{fig:fig4}(c)). Fig.~\ref{fig:fig4}(d)  plots $f_n(D, n/n_0=-2)$ versus $D/\epsilon_{0}$ over the temperature range $60$~$\mathrm{K}$-- $100$~$\mathrm{K}$. These data points collapse on top of each other with $f_n$ growing quadratically with $D/\epsilon_{0}$.

Note that $f_n$ has a slight asymmetry under sign-reversal of $D/\epsilon_{0}$. To understand this, we recall that the sign of layer polarization in BLG depends on the direction of $D$.  A positive $D$--field (as defined in Fig.~\ref{fig:fig1}(a)) increases the potential energy of electronic states in the lower layer of BLG as compared to those in the upper layer of BLG. For negative $n$, the occupied electronic states are mainly localized in the top layer of the BLG (that forms the moir\'e with the hBN)~\cite{PhysRevB.107.144111}. For the negative $D$--field, on the other hand,  the occupied electronic states are mainly localized in the bottom layer of the BLG (that does not form the moir\'e with the hBN). We postulate that the combined effect of this asymmetry of layer polarization on the sign of $D$ and the asymmetry of the moir\'e potential inherent in this device architecture ultimately manifests as $f_n(D) \neq f_n(-D)$.

With further increase in the displacement field,  $\alpha$ deviates from two, indicating a suppression of Umklapp processes for $|D/\epsilon_{0}|>0.3~\mathrm{V/nm}$. We do not have a clear understanding of the origin of this. One plausible reason can be that at large $D$, the trigonal warping becomes strong, severely limiting the phase space over which Eqn.~\ref{Eqn:Uee} may be satisfied~\cite{Seiler2022}. A related effect of the trigonal warping is the formation of overlapping electron-hole bands at certain number densities -- the scattering between thermally excited electrons and holes then masks Uee processes~\cite{PhysRevB.107.144111, Seiler2022}. A second possible cause of the suppression of Uee at high $D$ can be the strong modification of the BLG band by the displacement field (this includes layer-polarization, the opening of a band gap, and enhanced trigonal warping) leading to strong Zitterbewegung, which becomes the relevant scattering mechanism at large $|D|$~\cite{PhysRevB.87.115438}. Further experimental and theoretical studies are required to verify if any of these is indeed the cause for  suppression of Umklapp scattering with increasing~$D$.
\section{Conclusion}
To conclude, our experiments unequivocally establish Umklapp scattering to be the leading source of resistance in hBN/BLG superlattices in certain filling fraction ranges. Our findings on hBN/BLG superlattice differ from recent studies on hBN/SLG superlattice~\cite{Wallbank2019} in several significant aspects. In SLG hBN moir\'{e}, $\mathrm{\mathrm{R}}_{Uee}$ increases monotonically with increasing superlattice period and charge carrier density~\cite{Wallbank2019}. In contrast, $\mathrm{\mathrm{R}}_{Uee}$ in BLG moir\'e superlattice is predicted to have a non-monotonic dependence on superlattice period~\cite{PhysRevB.107.144111}. In this Letter, we have experimentally verified this prediction. Additionally, bilayer-based systems provide strong electric field tunability of the band gap and layer polarization and thus have an enormous scope for room-temperature applications ~\cite{tiwari2021electric, Zhang2009, https://doi.org/10.1002/aelm.202200510, PhysRevLett.105.166601, PhysRevB.106.205134, HE2020435}. We have shown that the strength of Uee increases rapidly with the increasing strength of the displacement field; this fact must be factored in when designing any $D$-field controlled superlattice device architectures. Additionally, we find the strength of Uee scattering to be stronger in BLG/hBN superlattice than in SLG/hBN superlattice (Supplementary Information, section S6).

With the presently available technology, the best quality BLG field effect devices are formed when encapsulated between a crystalline insulator, like hBN ~\cite{Uwanno2018, Hasan2021, Petrone2015}. As the growth of graphene in hBN leads to aligned layers ~\cite{Tang2013, Wang2021, Summerfield2016}, it is imperative to understand the significant sources of Joule heating in such systems for optimal room-temperature operations. Our present study achieves this and should motivate further studies in related systems like twisted bilayer graphene and twisted bilayers of transition metal dichalcogenides.

While this manuscript was under review, we became aware of a preprint ~\cite{shilov2023highmobility} which demonstrates that at $n/n_0 = -2$, transport in BLG/hBN moir\'e is dominated by Umklapp scattering.

\textbf{Acknowledgements:} A.B. acknowledges funding from U.S. Army DEVCOM Indo-Pacific (Project number: FA5209   22P0166) and Department of Science and Technology, Govt of India (DST/SJF/PSA-01/2016-17). M.J. and H.R.K. acknowledge the National Supercomputing Mission of the Department of Science and Technology, India, and the Science and Engineering Research Board of the Department of Science and Technology, India, for financial support under Grants No. DST/NSM/R$\&$D\_HPC Applications/2021/23 and No. SB/DF/005/2017, respectively. M.K.J. and R.B. acknowledge the funding from the Prime Minister's research fellowship (PMRF), MHRD. S.M. acknowledges the funding from the National post doctoral fellowship (N-PDF), SERB. K.W. and T.T. acknowledge support from the JSPS KAKENHI (Grant Numbers 21H05233 and 23H02052) and World Premier International Research Center Initiative (WPI), MEXT, Japan.

\textbf{Author contributions:} M.K.J., S.M., H.K.M., and A.B. conceived the idea of the study,  conducted the measurements, and analyzed the results. T.T. and K.W. provided the hBN crystals. R.B., M.J., and H.R.K. developed the theoretical model. All the authors contributed to preparing the manuscript.

\clearpage

\clearpage
	
	\section*{Supplementary Materials}
	
	\renewcommand{\theequation}{S\arabic{equation}}
	\renewcommand{\thesection}{S\arabic{section}}
	\renewcommand{\thefigure}{S\arabic{figure}}
	\renewcommand{\thetable}{S\arabic{table}}
	\setcounter{table}{0}
	\setcounter{figure}{0}
	\setcounter{equation}{0}
	\setcounter{section}{0}

	\section{\textbf{Device fabrication}}

\begin{figure*}[h]
	\includegraphics[width=0.8\columnwidth]{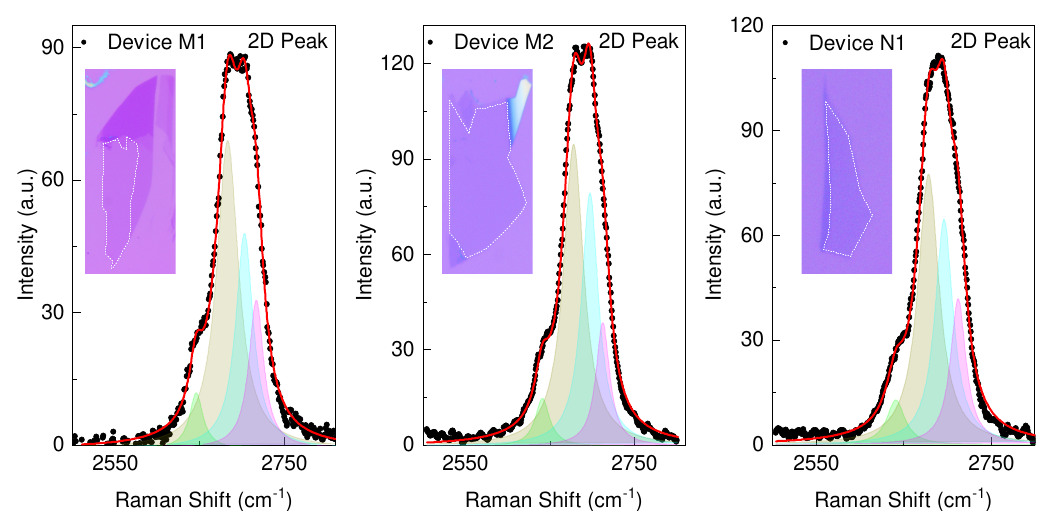}
	\small{\caption{ \textbf{Raman spectra of BLG flakes}. Plots of the $\mathrm{2D}$ Raman peak of the bilayer graphene used to fabricate devices (a) {M1}, (b) {M2}, and (c) {N1}. The black-filled circles are the experimentally measured Raman spectra.  The red solid line is cumulative of the four Lorentzian fitted to it; the four Lorentzian are also individually shown. The insets show the optical images of the  BLG flakes; the shaded region marks the bilayer graphene used in device fabrication.}
		\label{fig:figS1}}
\end{figure*}

The hBN encapsulated bilayer graphene (BLG) devices were fabricated using the dry transfer technique~\cite{doi:10.1021/acs.nanolett.3c00045, PhysRevLett.129.186802, Tiwari2022}. Initially, BLG and hBN flakes were mechanically exfoliated onto a Si/\ch{SiO2} substrate. BLG flakes were first identified with optical contrast and later confirmed with Raman spectra (Fig.~\ref{fig:figS1}). hBN flakes of thickness ranging from $25 - 30$~nm were used in fabrication, and their thickness uniformity was confirmed using an AFM.

To make BLG-hBN single-moir\'{e} devices (labeled {M1, M2, M3 and M4}), the top hBN sharp edge was aligned with the sharp edge of BLG at nearly zero degrees, facilitating a moir\'{e} between BLG and the top hBN layer. The bottom hBN edge was intentionally misaligned with BLG to prevent any moir\'{e} superlattice formation between BLG and the bottom hBN layer.

In the case of a non-moir\'{e} device ({N1}), hBN was misaligned with both the bottom hBN and top hBN layers to prevent any moir\'{e} superlattice formation. Electrical contacts were patterned through lithography, etched with a mixture of CHF$_3$ ($40$ sscm) and O$_2$ ($10$ sscm), and contacts deposited with Cr/Pd/Au with thickness $5/12/55$ nm to form $1$D contact with BLG. The device was etched in Hall bar geometry. Finally, The top gate was patterned through lithography, and a metal gate was deposited. Having dual gates in the devices gives control in tuning the system's carrier density and displacement field independently.

\section{\textbf{Twist angle estimation}}

\begin{figure}[h]
	\includegraphics[width=0.6\columnwidth]{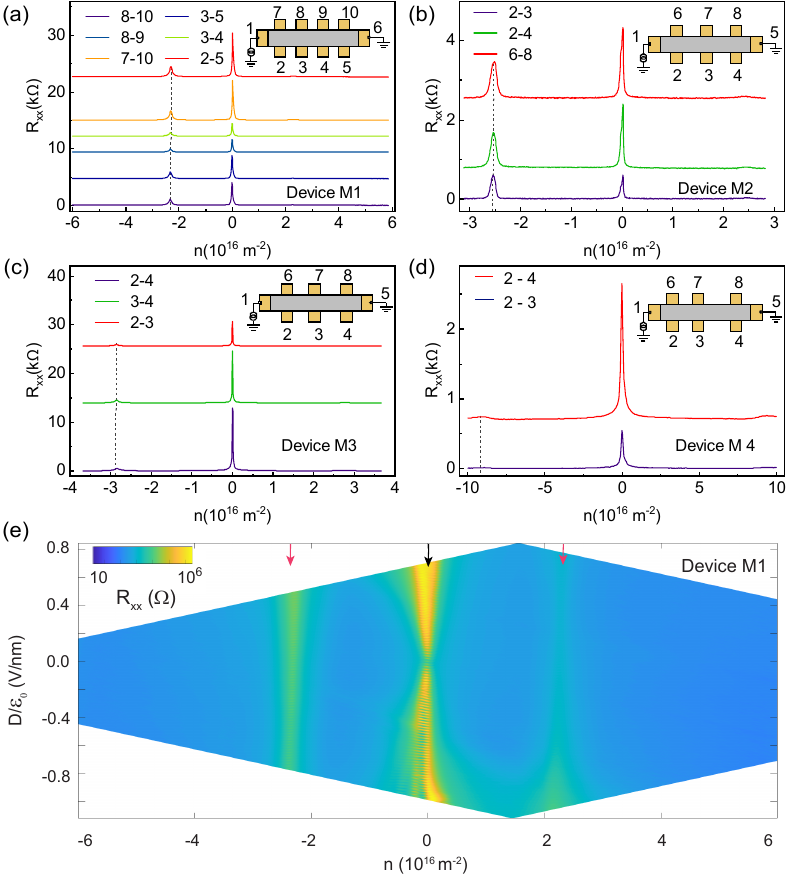}
	\small{\caption{ \textbf{Twist angle homogeneity of the moir\'{e} device.} (a) Plot of longitudinal resistance $\mathrm{R_{xx}}$ as a function of carrier density $n$ measured at  $T=2$~$\mathrm{K}$ for several configuration for  device {M1}. The vertical dashed line marks the carrier density at which the secondary moir\'{e} gap emerges.  The current was sourced between contacts 1 and 6. The numbers in the legend are the pairs of contacts used as voltage probes for the 4-probe measurement. [(b),(c),(d)] Same as in (a) for the devices {M2}, {M3}, and  {M4}, respectively. For four devices, the carrier density of the secondary moir\'{e} gap remains consistent across different configurations, illustrating the angle homogeneity within the devices. The plots are vertically offset for clarity. (e) 2D Plot of $\mathrm{R_{xx}}$ in the carrier density and the electric field $D/\epsilon_0$ plane. The magenta (black) arrows mark the position of the moir\'{e} gap (CNP), illustrating that the position of the moiré gaps is independent of the applied perpendicular electric field.}
		
		\label{fig:figS2}}
\end{figure}

Fig.~\ref{fig:figS2} shows the plot of longitudinal resistance R$_{xx}$ versus carrier density $n$ for devices {M1, M2, M3} and {M4} measured for different pairs of voltage leads at $T=2$~$\mathrm{K}$. The resistance peak at $n= 0$ originates from the charge neutrality point of the BLG. The resistance peak at $n_{M1}= -2.30\times 10^{16} ~\mathrm{m^{-2}}$, $n_{M2}= -2.49\times 10^{16} ~\mathrm{m^{-2}}$, $n_{M3}= -2.80\times 10^{16} ~\mathrm{m^{-2}}$ and $n_{M4}= -9.07\times 10^{16} ~\mathrm{m^{-2}}$, for device {M1}, {M2}, {M3} and {M4}, respectively, is a consequence of all levels up to the moir\'{e} gap getting filled at this carrier density. The independence of the carrier density at which the moir\'{e} gap emerges, regardless of the voltage probes used for the measurements (indicated with dotted lines), confirms the angle homogeneity in the devices. Further, Fig.~\ref{fig:figS2}(e) also illustrates that the carrier density corresponding to the moir\'{e} gap is independent of the applied displacement field.

\begin{figure*}[h]
	\includegraphics[width=0.75\columnwidth]{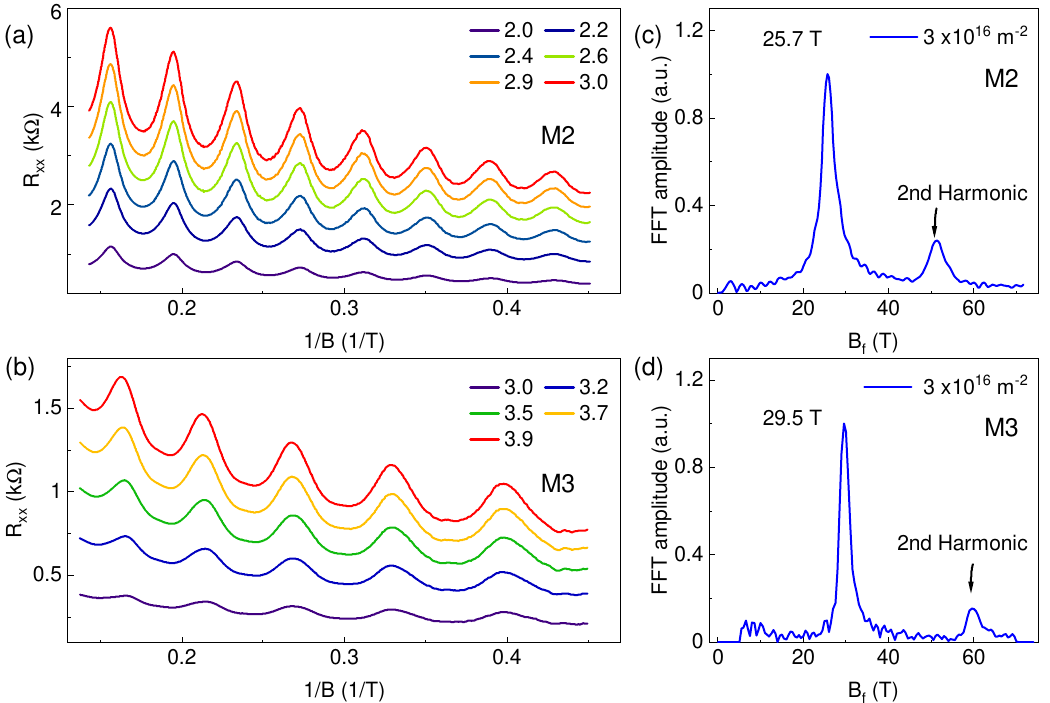}
	\small{\caption{ \textbf{ Brown-Zak Oscillations of the moir\'{e} device {M2} and {M3} .} (a) Plot of Brown-Zak oscillations $R_{xx}$  versus $1/B$ at a few representative values of $n$  measured at $T = 100$~$\mathrm{K}$, for device {M2}. The legends in the plot indicate the carrier density in units of $10^{16}$~$\mathrm{m^{-2}}$. The data are vertically offset for clarity. (b) Same as in (a) for device {M3}.  (c) The Fourier spectrum of the Brown-Zak oscillations measured at $n= 3\times10^{16}$~$\mathrm{m^{-2}}$, for device {M2}. The peak at $25.7$~T  corresponds to moir\'{e} wavelength of $13.64$~nm. (d) Same as in (c) for device {M3}, the peak at $29.5$~T  corresponds to moir\'{e} wavelength of $12.73$~nm.   }
		
		\label{fig:figS3}}
\end{figure*}

The moir\'{e} wavelength is estimated using the relation~\cite{doi:10.1126/sciadv.abd3655,jat2023higherorder,Sun2021}:
\begin{eqnarray}
	\lambda^2 = \frac{8}{\sqrt{3} n_M}  \label{Eqn:moirelength}
\end{eqnarray}
We find $\lambda_1=14$~nm for device {M1}, $\lambda_2=13.64$~nm  for device {M2}, $\lambda_3=12.73$~nm  for device {M3} and  $\lambda_4=7.20$~nm  for device {M4}.

We rule out dual-alignment of the BLG with both top - and bottom-hBN~\cite{jat2023higherorder,doi:10.1126/sciadv.aay8897} through measurements of the Brown-Zak oscillations of conductance at $T = 100$~$\mathrm{K}$. At these elevated temperatures, landau levels get smeared out, and only the magnetotransport oscillations from the recurring Bloch states in the superlattice survive, and are shown in  Fig.1(e) of the main text for device {M1} and in Fig.~\ref{fig:figS3}(a,b) for device {M2, M3}. The periodicity of these oscillations is independent of the carrier density. The fast Fourier transform of these oscillations yields a single ``frequency'' $B_f=24.2$~T, $25.7$~T and $29.5$~T for device {M1, M2} and {M3} respectively, limiting the possibility of supermoir\'{e} in the system.  We note that a single Brown-Zak oscillation frequency can also occur if the two twist angles between the top hBN and BLG and that between the BLG and bottom hBN are identical; given that we intentionally misaligned the bottom hBN by a large angle, we rule out this scenario.

The frequency $B_f$ is related to the real-space area $S$ of the moir\'{e} unit cell by $B_f =\phi_0/S$, where $\phi_0 = h/e$ is the flux quantum~\cite{PhysRevB.14.2239,doi:10.1126/sciadv.abd3655,Huber2022,doi:10.1126/science.aal3357}. {Using the relation $\lambda = \sqrt{(2S/\sqrt{3})}$}, we estimate $\lambda_1=14$~nm (for device {M1}), $\lambda_2=13.64$~nm (for device {M2}) and $\lambda_3=12.73$~nm (for device {M3}). These values match exactly with the moir\'{e} wavelength extracted from resistance versus carrier density response.

We estimate the twist angle magnitude between BLG and hBN using the relation~\cite{doi:10.1126/science.1237240, doi:10.1021/acs.nanolett.8b05061}:
\begin{eqnarray}
	\mathrm{\lambda = \frac{(1+\epsilon)a}{[\epsilon^2 +2(1+\epsilon)(1-cos(\theta))]^{1/2}}}\label{Eqn:moireangle}
\end{eqnarray}
Here $\mathrm{a} = 0.246~\mathrm{nm}$ is the lattice constant of graphene, $\epsilon =0.018$ is the lattice mismatch between the hBN and graphene, and $\theta$ is the relative twist angle between hBN and BLG. We find the twist angle between the BLG and hBN to be $\theta_{M1} = 0^\circ$, $\theta_{M2} = 0.26^\circ$, $\theta_{M3} = 0.47^\circ$ and $\theta_{M4} = 1.70^\circ$ for device {M1, M2, M3 and M4}, respectively.

\section{\textbf{Comparison of sheet resistance  between moir\'{e} and non moir\'{e} device}}

\begin{figure}[h]
	\includegraphics[width=0.75\columnwidth]{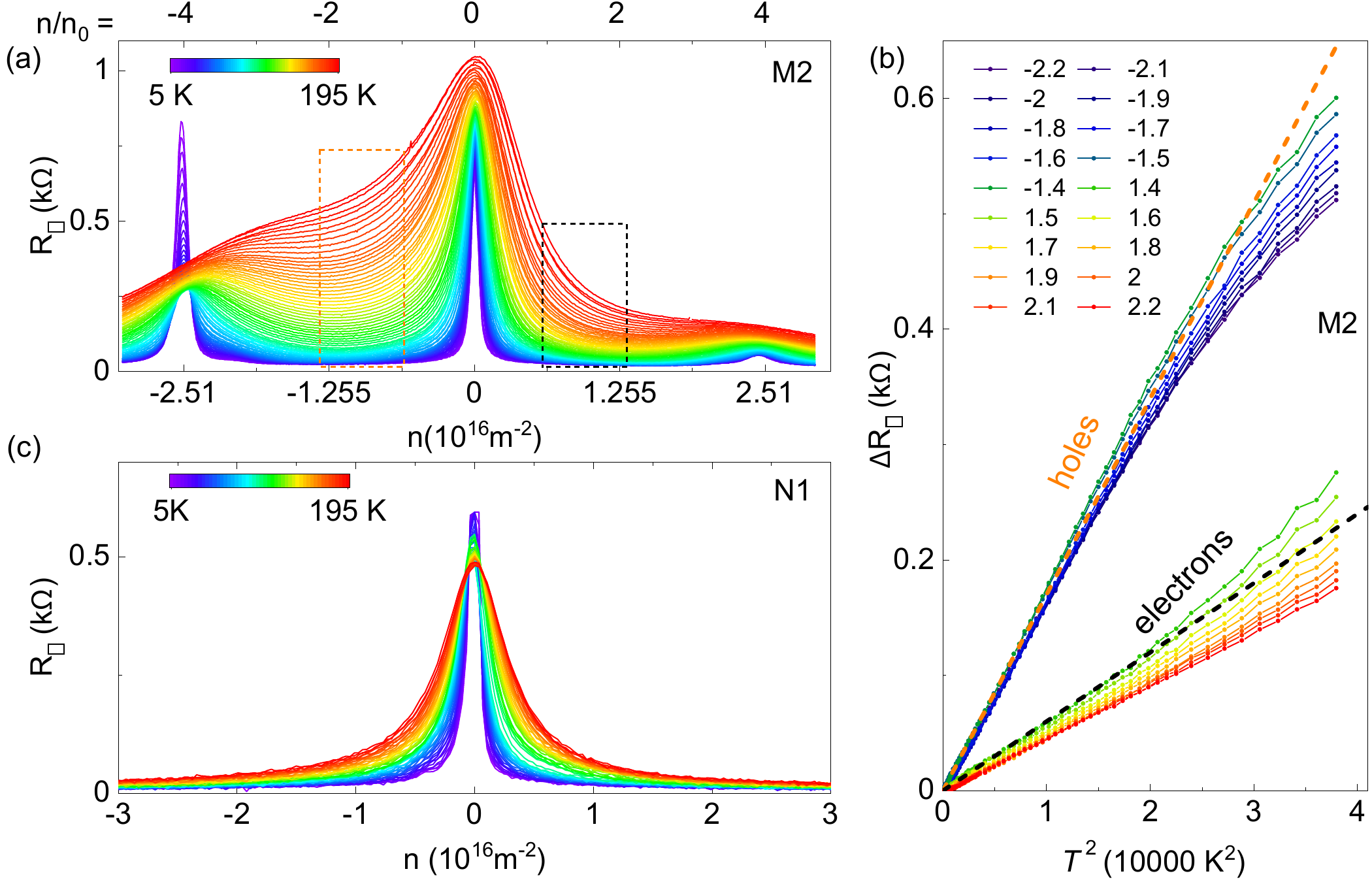}
	\small{\caption{ \textbf{Electron-Hole asymmetry of the moir\'{e} device and comparison with non-moir\'e device.} (a) Plot of longitudinal sheet resistance $\mathrm{R_\square}$ as a function of filling fraction $n/n_0$ over a range of $T$ from $5$~$\mathrm{K}$ (blue) to $195$~$\mathrm{K}$ (red). The measurements were done with $B=0$ and $D=0$. The dotted rectangles mark the regions where Umklapp is the dominant scattering mechanism. The plot illustrates the particle-hole asymmetry in the strength of Uee. (b) Plots of $\mathrm{\Delta R_\square}$ versus $T^2$ at a few representative filling fractions $(n/n_0)$ for electron and hole doping.  The numbers in the legend are the values of $n/n_0$. The dotted lines are linear fits to the data at $n/n_0= \pm 1.4$. (c) $T$-dependence of $\mathrm{R_\square}$ for the non-aligned device {N1} -- note the absence of the large resistance enhancement with $T$, as seen in the case of the moir\'{e} devices.}
		
		\label{fig:figS4}}
\end{figure}

Fig.~\ref{fig:figS4}(a) shows the longitudinal sheet resistance versus carrier density response for device {M2} over a range of temperatures. The Umklapp electron-electron scattering dominated region is marked with dotted rectangles. The hole side (marked with an orange rectangle) shows a significantly larger strength of umklapp strength than the electron side (marked with a black rectangle). Fig.\ref{fig:figS4}(b) plots the sheet resistance versus $T^2$ to better show this electron-hole asymmetry over a range of $n/n_0$. The dotted orange (black) line shows a guiding straight line for holes (electrons) type carriers. This large asymmetric strength origin can be attributed to the much larger probability of the back-scattering of holes than that of electrons~\cite{Wallbank2019}.

The $T$-dependence of the sheet resistance in non-aligned device {N1} are plotted in Fig.~\ref{fig:figS4}(c) for comparison. The data in the non-aligned device differ from that of {M1} and {M2} in three important aspects:
\begin{enumerate}
	\item The large increase in sheet resistance seen in {M1} and {M2} with increasing $T$ is conspicuously absent in the non-aligned device {N1}.
	\item A $T^2$-dependence of the sheet resistance is not observed for the device {N1} (Fig.2(d) of the main manuscript). This is expected since, in a non-aligned device, Umklapp scattering is forbidden ~\cite{PhysRevB.107.144111}.
	\item In contrast to that of {M1} and {M2}, the electron-hole asymmetry in sheet resistance is absent for device {N1}.
\end{enumerate}

\section{\textbf{Calculation of threshold density}}

\begin{figure}[h]
	\includegraphics[width=0.75\columnwidth]{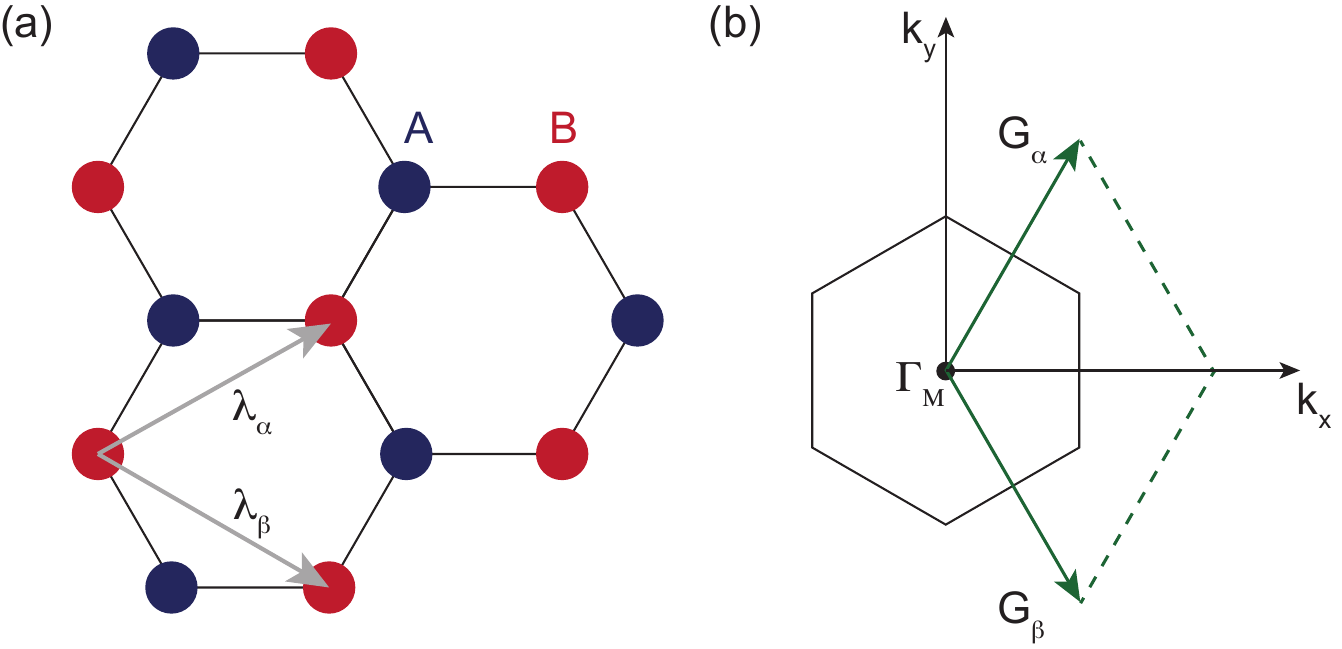}
	\small{\caption{ \textbf{Moir\'{e} lattice and its Brillouin zone.} (a) Real space lattice of the moir\'{e} lattice, $\mathbf{\lambda_\alpha}$ and $\mathbf{\lambda_\beta}$ are lattice unit vectors. (b) Brillouin zone with reciprocal lattice vectors $\mathbf{G_\alpha}$ and $\mathbf{G_\beta}$.}
		
		\label{fig:figS5}}
\end{figure}

The real-space lattice of bilayer graphene and hBN leads to a hexagonal moir\'{e} lattice (Fig.~\ref{fig:figS5}(a)). The lattice vectors can be written as:
\begin{align}
	&\mathbf{\lambda_\alpha} = \frac{\lambda}{2}( \sqrt{3}, 1),   &\mathbf{\lambda_\beta} = \frac{\lambda}{2}( \sqrt{3}, -1).
\end{align}
where  $\lambda$ is the moir\'{e} wavelength.
The corresponding reciprocal lattice vectors are given by (Fig.\ref{fig:figS5}(b)):
\begin{align}
	&\mathbf{G_\alpha} = \frac{2\pi}{\sqrt{3}\lambda}(1, \sqrt{3}),   &\mathbf{G_\beta} = \frac{2\pi}{\sqrt{3}\lambda}(1, -\sqrt{3}).
\end{align}

In the Uee scattering process, the condition for backscattering is
\begin{eqnarray}
	\mathbf{k_1 + k_2 = k_3 + k_4 + G} \label{Eqn:Ueeb}
\end{eqnarray}
here, $\mathbf{k_1, k_2}$ are the wave-vectors of the incoming electrons, $\mathbf{k_3, k_4}$ are the wave-vectors of the  scattered electrons and $\mathbf{G}$ is a reciprocal space lattice vector.

The lower limit on wave-vector ${k_t}$  above which umklapp scattering is allowed comes from Eqn.~\ref{Eqn:Ueeb}, which gives ${k_t}= {G}/4 = \pi/({\sqrt{3}\lambda})$. The
corresponding threshold carrier density above which Umklapp scattering starts can be written as $n_t =\mathbf{|k_t|}^2/\pi=\pi/({3\lambda^2})= n_0\pi/({2\sqrt{3}}) \approx 0.907 n_0$. (assuming an isotropic dispersion). Here, $n_0 = 2/({\sqrt{3}\lambda^2})$ is the carrier density corresponding to one-fourth filling of moir\'{e} band.

\section{\textbf{Quantum Hall of the Device}}

\begin{figure}[h]
	\includegraphics[width=0.75\columnwidth]{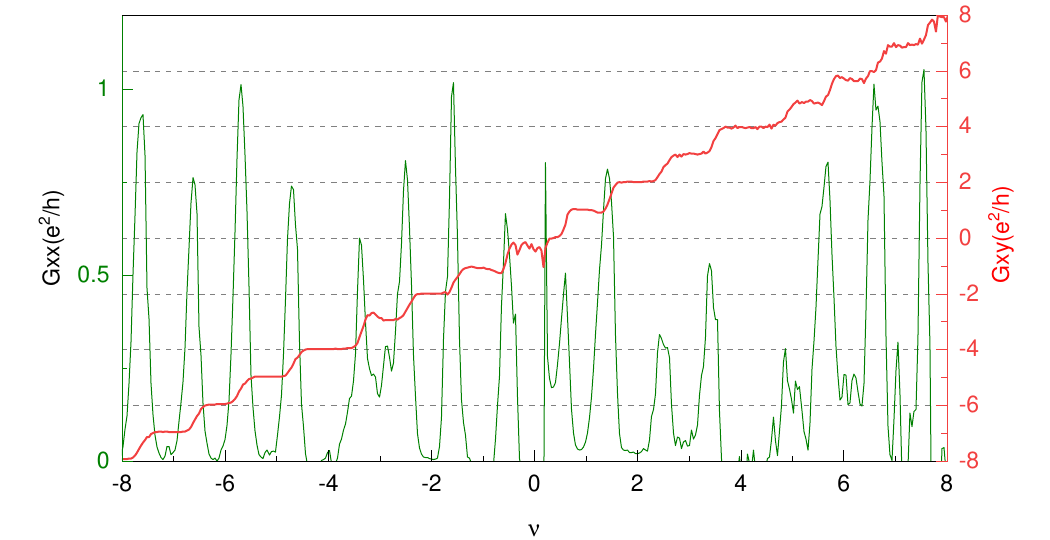}
	\small{\caption{ \textbf{Quantum Hall of the device.} Plots of $\mathrm{G_{xx}}$ (green solid line) and $\mathrm{G_{xy}}$ (red solid line) versus the filling factor $\nu= nh/eB$. The measurements were done for $B = 5$~T and  at $T=20$~mK.}
		
		\label{fig:figs6}}
\end{figure}

The Quantum Hall measurements at a perpendicular magnetic field of $B=5$~T were performed to estimate the value of capacitance $C_{bg}$  and $C_{tg}$. Fig.\ref{fig:figs6} shows $\mathrm{G_{xx}}$ (green solid line) and $\mathrm{G_{xy}}$ (red solid line) versus the filling fraction ($\nu= nh/eB$)  measured at a perpendicular magnetic field of $5$~T,  establishing that both spin and valley degeneracies are lifted, indicating the high quality of the device.

\section{\textbf{Comparision of Umklapp strength in the hBN moir\'{e} of BLG and SLG}}
Fig.\ref{fig:figs7} plots the $f_n$ versus moir\'{e} wavelength for the hBN moir\'{e}  devices, measured at $n/n_0 = -2$. The Red data points are from our measured BLG-hBN moir\'{e} devices, and the blue data points of the SLG-hBN moir\'{e} devices from Ref~\cite{Wallbank2019}. As a similarity, both systems show the Umklapp scattering at $n/n_0 = -2$.

However, there are significant differences between these two material systems ~({Fig.\ref{fig:figs7}):
	\begin{enumerate}
		\item The strength of $f_n$ is larger in BLG-hBN moir\'{e}  than in the SLG-hBN moir\'{e}.
		\item  $f_n$ depends non-monotonically on the moir\'{e} wavelength for BLG-hBN moir\'{e}, whereas for SLG-hBN moir\'{e}, $f_n$ increases monotonically with the moir\'{e} wavelength.
	\end{enumerate}
	
	\begin{figure}[t]
		\includegraphics[width=0.7\columnwidth]{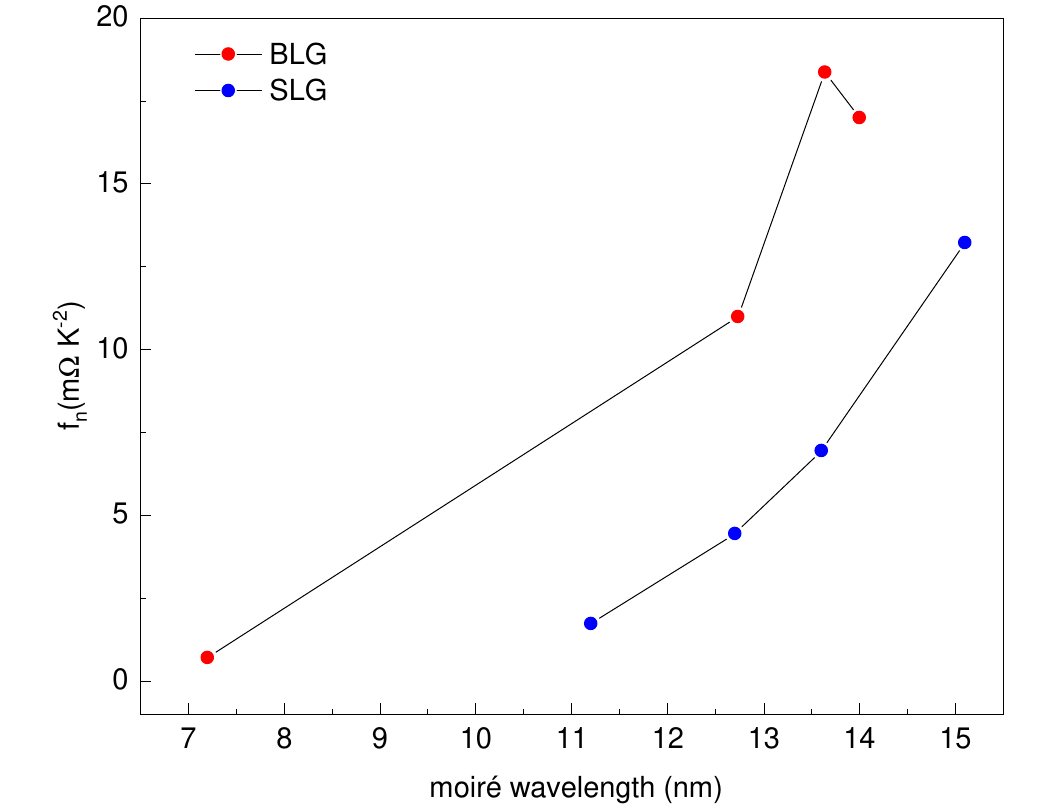}
		\small{\caption{ \textbf{Umklapp strength in the hBN moir\'{e} of BLG and SLG.} Plot of $f_n$ versus moir\'{e} wavelength, measured at $n/n_0 = -2$ . Red data points are from our measured BLG-hBN moir\'{e} devices, and blue data points are of the SLG-hBN moir\'{e} devices taken from Ref~\cite{Wallbank2019}}.
			
			\label{fig:figs7}}
	\end{figure}
	
	\clearpage

\end{document}